\documentclass[pra,twocolumn,showpacs,superscriptaddress]{revtex4}
\usepackage{graphicx,color,graphics}
\usepackage{amssymb,amsmath,verbatim,ulem}

\begin{document}

\title{Long-range 1D gravitational-like interaction in a neutral atomic cold gas}%
\author{M. Chalony}
\affiliation{Institut Non Lin\'eaire de Nice, Universit\'e de Nice Sophia-Antipolis, CNRS, 06560 Valbonne, France.}
\author{J. Barr\'e}
\email{Julien.Barre@unice.fr}
\affiliation{Laboratoire J.-A. Dieudonn\'e, Universit\'e de Nice Sophia-Antipolis, CNRS, 06109 Nice, France.}
\author{B. Marcos}
\email{bruno.marcos@unice.fr}
\affiliation{Laboratoire J.-A. Dieudonn\'e, Universit\'e de Nice Sophia-Antipolis, CNRS, 06109 Nice, France.}
\author{A. Olivetti}
\affiliation{Laboratoire J.-A. Dieudonn\'e, Universit\'e de Nice Sophia-Antipolis, CNRS, 06109 Nice, France.}
\author {D. Wilkowski}
\email{david.wilkowski@ntu.edu.sg}
\affiliation{Institut Non Lin\'eaire de Nice, Universit\'e de Nice Sophia-Antipolis, CNRS, 06560 Valbonne, France.}
\affiliation{Centre for Quantum Technologies, National University of Singapore, 117543 Singapore, Singapore.}
\affiliation{PAP, School of Physical and Mathematical Sciences, Nanyang Technological University, 637371 Singapore, Singapore.}
\date{\today{}}
\begin{abstract}
  A quasi-resonant laser induces a long-range attractive force within
  a cloud of cold atoms. We take advantage of this force to build in
  the laboratory a system of particles with a one-dimensional gravitational-like
  interaction, at a fluid level of modeling.  We give experimental
  evidences of such an interaction in a cold Strontium gas, studying
  the density profile of the cloud, its size as a function of the
  number of atoms, and its breathing oscillations.
\end{abstract}
%

\maketitle
\section{Introduction}

When interactions between the microscopic components of a system act
on a length scale comparable to the size of the system, one may call
them ``long-range'':  For instance, the inverse-square law of the
gravitational force between two point masses which is one of the most
celebrated and oldest laws in physics. In the many particles world, it
is responsible for dramatic collective effects such as the
gravothermal catastrophe~\cite{antonov1961} or the gravitational
clustering which is the main mechanism leading to the formation of the
structure of galaxies in the present universe.  Beyond gravitation,
such long-range interactions are present in various physical fields,
either as fundamental or as effective interactions: In plasma physics \cite{refplasma},
two dimensional (2D) fluid dynamics \cite{reffluid}, degenerated quantum gases \cite{PhysRevLett.84.5687}, ion
trapping \cite{refion}, to cite only these works. long-range interactions deeply influence the dynamical and thermodynamical
properties of such systems.  At the thermodynamic equilibrium, long-range interactions are at the origin of very peculiar properties,
especially for attractive systems: The specific heat may be negative;
canonical (fixed temperature) and microcanonical (fixed energy)
ensembles are not equivalent.  These special features have been known
for a long time in the astrophysics community, in the context of self
gravitating systems.

After the seminal works of Lynden-Bell and Wood \cite{LyndenBell68}
and Thirring \cite{Thirring70}, many contributions followed on this
subject (see, for instance, \cite{Review_Longrange2009} for a recent
review), so that the equilibrium characteristics of attractive long-range interacting systems are theoretically well established. This
situation is in striking contrast with the experimental side of the
problem: There is currently no controllable experimental system
exhibiting the predicted peculiarities.  There have been some
proposals to remedy to this situation: O'Dell \emph{et} \emph{al.}
\cite{PhysRevLett.84.5687} have suggested creating an effective $1/r$
potential between atoms in a Bose-Einstein condensate using
off-resonant laser beams; more recently, Dominguez \emph{et}
\emph{al.}~\cite{dominguez2010dynamics} have proposed taking
advantage of the capillary interactions between colloids to mimic
two-dimensional gravity, and Golestanian~\cite{Golestanian} has
suggested experiments using thermally driven colloids. However, these
proposals have not been implemented yet, and so far the dream of a
tabletop galaxy remains elusive.

The key results of this paper are to show some
  experimental evidences of a gravitational-like interaction in a
  quasi-one-dimensional (hereafter 1D) test system consisting in a
  cold gas of Strontium atoms in interaction with two
  contra-propagating quasi-resonant lasers. To our knowledge, it is the first experimental realization of the 1D
  gravitational toy model, which can be compared
  with the theoretical predictions developed for more than $50$ years
  by the astrophysical and statistical physics community. In the stationary regime,
  the cloud spatial distribution is in agreement with the well-known
  $\mbox{sech}^2$ law for the 1D self-gravitating gas at thermal
  equilibrium \cite{camm50self}. Moreover, the long-range attractive
  nature of the force is confirmed studying the cloud's size
  dependency as a function of the number of atoms. Out of equilibrium,
  the breathing oscillation frequency increases with the strength of
  the interaction as it should be for attractive
  interactions. Quantitatively our experimental results are in
  agreement with the expected $1/r^{\alpha}$ force with $\alpha=0$.

The paper is organized as follows. In Sec.~\ref{sec:model}, we
start from the radiation pressure exerted by the lasers and explain
under which circumstances this force becomes analog to a 1D
gravitational force. We then make some definite theoretical
predictions on the size, density profile, and oscillation frequency of
the interacting atomic cloud. The experimental setup is described in
Sec.~\ref{sec:experiments}. In the same section, the experimental results are
compared with the theory.

\section{Model and theoretical predictions}
\label{sec:model}
The gravitational potential $U(r)$ between two particles can be
expressed through the Poisson equation $\nabla^2 U(\mathbf{r}) = A_D G
m \delta(\mathbf{r})$, where $G$ is the coupling constant, $m$ the
mass of the particle and $A_D$ a numerical constant which depends on
the dimension. The solution of the Poisson equation for the
interpaticle potential $U(r)$ in three dimension is the well-known
\begin{equation}
U(r)=\frac{Gm}{r},
\end{equation}
and in 1D,
\begin{equation}
\label{grav_1D}
U(r)=Gm|r|
\end{equation}
(for a review on 1D gravitational systems see,
e.g., \cite{Miller}). After using a mean-field approach (see below), we will show that such a potential should be at
play in our experiment, under precise circumstances (see section \ref{subsec:stationary_solution}).

We start considering a quasi-1D $\{$cold atomic gas + 1D
quasi-resonant laser beams$\}$ system; an atomic gas,
with a linear density $n(z)$, is in interaction with two
contra-propagating laser beams. The two beam intensities $I_+(z)$
and $I_-(z)$, where $I_+(-\infty)=I_-(+\infty)\equiv I_0$,
respectively propagating in the positive and negative direction, are
much smaller than the atomic line saturation intensity $I_s$. Thus the
atomic dipolar response is linear. The radiation pressure force of the
lasers on an atom, having a longitudinal velocity $v_z$, is given
by \cite{metcalf1994cooling}:
\begin{equation}
  F_\pm(z,v_z)=\pm \hbar
  k\frac{\Gamma}{2}\frac{\Gamma^2}{4(\delta\mp kv_z)^2+\Gamma^2}\frac{I_\pm(z)}{I_s},
\label{eq_rad_F}
\end{equation}
where $\hbar$ is the reduced Planck constant, $\Gamma$ the bare
linewidth of the atomic transition, $k$ the wave number, and $\delta$
the frequency detuning between an atom at rest and the lasers.
For a cloud of $N$ atoms, the attenuation of the
laser intensity is given by:
\begin{equation}
\textrm{d}I_\pm=\mp\frac{\sigma_{\pm}}{2\pi L_{\perp}^2} NI_\pm n(z)\textrm{d}z, \label{eq_dI}
\end{equation}
 where $n(z)$ is the normalized linear density profile and
\begin{equation}
\sigma_{\pm}=\frac{6\pi}{k^2}\Gamma^2\int\frac{g(v_z)}{4(\delta\mp kv_z)^2+\Gamma^2}\textrm{d}v_z
\label{eq_sigma}
\end{equation}
is the average absorption cross-section for a single atom. $g(v_z)$ is the normalized longitudinal velocity distribution and $2\pi
L_{\perp}^2$ is the transverse section of the cloud. At equilibrium
$g(v_z)$ is an even function so $\sigma_-=\sigma_+\equiv\sigma$. The optical depth is defined as:
\begin{equation}
  b=\frac{\sigma}{2\pi  L_{\perp}^2}N\int_{-\infty}^{+\infty}n(z)\textrm{d}z=\frac{\sigma N}{2\pi  L_{\perp}^2}\,.
\label{eq_b}
\end{equation}
Atoms also experience a velocity diffusion due to the random photon
absorptions and spontaneous emissions: This is modeled
by a velocity diffusion coefficient $D$ introduced in Eq.~(\ref{eq:VFP}). In experiments, $\delta<0$ such
that the force, given in Eq.~(\ref{eq_rad_F}), is a cooling force
counteracting the velocity diffusion.
We now describe the $N$ atoms by their phase space
density in 1D,
$f(z,v_z,t)$. As in \cite{Verkerk11}, we write a
Vlasov Fokker-Planck equation
\begin{equation}
\begin{split}
  \frac{\partial f}{\partial t}+v_z\frac{\partial f}{\partial z} &
-\omega_z^2z\frac{\partial f}{\partial v_z}
+\frac1m\frac{\partial}{\partial v_z}[(F_{+}(z,v_z)+F_{-}(z,v_z))f] \\
&=D\frac{\partial^2 f}{\partial v_z^2}.
\end{split}
\label{eq:VFP}
\end{equation}
which is, for most of the cases, a reasonable modeling of long-range force systems in the mean-field approximation (see, e.g., \cite{binney}).
The second term in Eq.~(\ref{eq:VFP}) is an inertial one, whereas the third
one describes a harmonic trapping force being a good approximation
of the dipolar trap used in the experiment \cite{chu86}. Indeed the dipolar potential, in the longitudinal axe of interest,
can be written as
\begin{equation}
U_{dip}(z)=\frac{-U_0}{1+\left(\frac{z}{z_R}\right)^2}
\label{eq:dipole_trap}
\end{equation}
with $z_R=1.2$ mm, $U_0=\frac12 k_BT_{\mathrm{trap}}$, and
$T_{\mathrm{trap}}=20\mu K$. The observed \emph {rms} longitudinal
size being $L_z\lesssim 400 \mu m$, it is reasonable to perform a
Taylor expansion around $z=0$ to get the harmonic approximation:
\begin{equation}
U_{dip}(z)\approx -U_0\left[1-\left(\frac{z}{z_R}\right)^2\right].
\end{equation}
having a characteristic frequency
\begin{equation}
\label{om-zr}
\omega_z=\left(\frac{k_B T_{\mathrm{trap}}}{m z_R^2}\right)^{1/2}.
\end{equation}
The fourth term of Eq.~(\ref{eq:VFP}) contains the mean-field force
$F_{\pm}$ divided by the atomic mass $m$.
The right hand side describes a velocity
diffusion.  The use of a one dimensional model is justified by the
fact that the ratio between the \emph{rms} transverse $L_\perp$ and longitudinal
$L_z$ size of the cloud measured in the experiment is
$L_\perp/L_z\approx 2\times 10^{-2}$. Eq.~\eqref{eq:VFP}
neglects atomic losses and dependencies in position and velocity of
the velocity diffusion coefficient.

One notes that the attractive force coming from the
  beams absorption (Eq.~\eqref{eq_rad_F} and \eqref{eq_dI}) is known
  since the early days of laser cooling and trapping
  \cite{dalibard1988laser}. However, in an usual 3D setting this
  attractive force is dominated by the repulsive force due to
  photons reabsorption \cite{walker1990collective}, which, in the small
  optical depth limit, may be seen as an effective repulsive Coulomb
  force. By contrast, in a 1D configuration with an elongated cloud
  along the cooling laser beams, the probability of photons
  reabsorption is reduced by a factor of the order of $L_\perp/L_z$,
  in comparison with the isotropic cloud having the same longitudinal
  optical depth.  In our experiment, the reduction factor is about
  $2\times 10^{-2}$, so that the repulsive force can be safely
  ignored. Similar but weaker reduction of the probability of photons
  reabsorption is also expected for the 2D geometry, which opens the
  possibility of experimental systems analogous to 2D self-gravitating
  systems.

\subsection{Fluid approximation}

In order to solve Eq.~\eqref{eq:VFP} we assume that the system can be
described using a fluid approach: The velocity
distribution at time $t$ does not depend on the position, except for a
macroscopic velocity $u(z,t)$. We write then the one point
distribution function $f$ as
\begin{equation}
f(z,v_z,t) = m N n(z,t) \frac{1}{\Delta(t)}g\left(\frac{v_z-u(z,t)}{\Delta(t)}\right).
\label{eq:f}
\end{equation}
The velocity distribution $g(v_z)$ is even,
centered around $u$;
the velocity dispersion is characterized by a time modulation
$\Delta(t)$. Integrating Eq.~(\ref{eq:VFP}) over $dv_z$ and over $v_z~dv_z$, we
obtain the fluid equations:
\begin{eqnarray}
\frac{\partial n}{\partial t} +\frac{\partial}{\partial z}\left( n u\right)
=  0 \quad &&\label{eq:fluid1}\\
\frac{\partial (n u)}{\partial t} +\frac{\partial}{\partial z}\left[\left( u^2 + \Delta(t)^2 \int v_z^2g(v_z)dv_z\right)n\right]
+\omega_z^2zn && \nonumber \\
-\frac1m n\int (F_++F_-)g\left(\frac{v_z-u(z,t)}{\Delta(t)}\right){\Delta(t)} dv_z = 0. \quad && \label{eq:fluid2}
\end{eqnarray}

\subsection{Stationary solution}
\label{subsec:stationary_solution}

We first look for a stationary solution; this imposes $u=0$ and $\Delta=1$.
Eq.~(\ref{eq:fluid1}) is then automatically satisfied;
Eq.~(\ref{eq:fluid2}) for the stationary density $n(z)$ reads:
\begin{eqnarray}
 \bar{v}_z^2\frac{\partial n}{\partial z} +\omega_z^2z n
- \frac1m n\int [F_{+}+F_{-}]g(v_z)dv_z &=&0\,,
\label{eq:stat1}
\end{eqnarray}
where we have used the notation $\int v_z^2g(v_z) dv_z = \bar{v}_z^2$.

Eq.~(\ref{eq_dI}) is easily integrated, yielding:
\begin{eqnarray}
I_+(z) &=&I_0e^{-b \int_{-\infty}^z n(s)ds} \\
I_-(z) &=&I_0e^{-b \int_z^{+\infty} n(s) ds}
\end{eqnarray}
The exponentials are expanded up to first order, according to the
small optical depth hypothesis $b\ll 1$:
\begin{eqnarray}
I_+(z)&\simeq& I_0\left(1-b \int_{-\infty}^z n(s)
  ds\right) \\
I_-(z)&\simeq& I_0\left(1-b \int_z^{+\infty} n(s)
  ds\right)~.
\end{eqnarray}
Introducing these expressions for $I_\pm$ into Eq.~(\ref{eq:stat1}), we
obtain finally
\begin{equation}
  \bar{v}_z^2\frac{\partial n}{\partial z} +\omega_z^2zn-NC n\int_{-\infty}^{+\infty}\mbox{sgn}(s-z)n(s)ds
  =0\,,\label{eq:stat2}
\end{equation}
where
\begin{equation}
C=\frac{3\hbar\Gamma}{2mkL_{\perp}^2}\frac{I_0}{I_s}\left(\sigma\frac{k^2}{6\pi}\right)^2~.
\end{equation}
Eq.~(\ref{eq:stat2}) is equivalent to an equation describing the
stationary density of an assembly of $N$ trapped
particles of mass $m$, with gravitational coupling constant $G$, in an
external harmonic trap of frequency $\omega_z$, in a heat bath at
temperature $T$, with the correspondence:
\begin{subequations}
\begin{align}
\label{corr1}
\bar{v}_z^2&\leftrightarrow \frac{k_BT}{m}\\
\label{corr2}
 C&\leftrightarrow Gm~,
\end{align}
\end{subequations}
where $k_B$ is the Boltzmann constant.
Two characteristic
lengths are identified;
\begin{equation}
\label{L_ni}
L_{ni}=\sqrt{\frac{k_B T}{m\omega_z^2}}
\end{equation}
is the characteristic size of the
non-interacting gas in its external harmonic holding potential.
Using Eq.~\eqref{om-zr} we get
\begin{equation}
L_{ni}=\sqrt{\frac{T}{T_{trap}}}z_R.
\end{equation}
The other characteristic length $L_i$ is associated with the
interaction strength:
\begin{equation}
L_i=\frac{k_BT}{NCm}.
\end{equation}
Using these notations we write Eq.~\eqref{eq:stat2} as
\begin{equation}
\label{eq:stat}
 \frac{\partial n}{\partial z} +\frac{zn}{L_{ni}^2}-\frac{n}{L_i}\int_{-\infty}^{+\infty}\mbox{sgn}(s-z)n(s)ds=0~.
\end{equation}
The first term of~\eqref{eq:stat} favors the density spreading. In
contrast with the 2D and 3D cases, it always prevents the collapse of
the cloud \cite{Padmanabhan}. The second term describes an
external harmonic confinement coming from the dipole trap in the
experiment. The third term is the attractive interaction due to laser beams absorption. It corresponds to an 1D
gravitational potential expresses in Eq. \eqref{grav_1D}. If the inequality
$L_i\ll L_{ni}$ is fulfilled, Eq. (\ref{eq:stat}) is the one expected
for a 1D self-gravitating gas at thermal
equilibrium~\cite{camm50self}. It yields the profile:
\begin{equation}
n(z)=\frac{1}{4L_i}\textrm{sech}^2\left(\frac{z}{2L_i}\right).
\label{eq:sech}
\end{equation}
A generalization of Eq.~(\ref{eq:stat}) is written as:
\begin{equation}
 \frac{\partial n}{\partial z} +\frac{1}{k_BT}
\frac{\partial U_{dip}}{\partial z}n-An\int_{-\infty}^{+\infty}|s-z|^{-\alpha}
\mbox{sgn}(s-z)n(s)ds=0~,
\label{eq:stat_gen}
\end{equation}
including the exact form of the dipole trap
(\ref{eq:dipole_trap}), and the variation of the interaction exponent
$\alpha$ of a $1/r^\alpha$ attractive force. This expression is used to compare theory with experiments in section \ref{sec:experiments}. $A$ is a free
parameter controlling the interaction strength, and thus the width of
the equilibrium profile.

\subsection{Breathing oscillations}
To probe the dynamics of the system, we now go back to
Eqs.~(\ref{eq:fluid1},\ref{eq:fluid2}), linearizing these equations with
respect to $u$ and $\Delta-1$: For small amplitude oscillations. One notes that this
approximation is much less restrictive than linearizing with respect to the
velocity $v_z$. We then compute $\int [F_++F_-]f dv_z$:
\begin{equation}
\begin{split}
\int [F_++F_-]f dv_z \simeq &~c_1(I_+-I_-)n +c_2(I_++I_-)n u \\
&+c_3(\Delta-1)(I_+-I_-)n~,
\end{split}
\end{equation}
where the constants $c_i$ involve integrations with respect to $v_z$.
The first term is the gravitational-like force, as in (\ref{eq:stat2})
with $n(z)$ replaced by the time-dependent density $n(z,t)$.  The
second one is a friction, which a priori depends weakly on $z$ through
$I_++I_-$. Since $I_+-I_-$ is of order $b\ll 1$, the third term, of
order $b(\Delta-1)$, is neglected. We assume that
the dynamics is captured by a single parameter $\lambda(t)$, using the ansatz \cite{PhysRevLett.103.224301}:
\begin{equation}
f(z,v_z,t)=mN n(z/\lambda) g(\lambda v_z -\dot{\lambda} z).
\label{eq:ansatz}
\end{equation}
When compared with \eqref{eq:f}, this amounts to assume:
$u=\frac{\dot{\lambda}}{\lambda}z,~\Delta=1/\lambda$.  We introduce
the notations $\langle.\rangle$ and $\langle.\rangle_0$ for the
spatial average of a quantity over the density at time $t$ and the
stationary density respectively.  Then
\begin{equation}
\langle z^2\rangle =\lambda^2\langle z^2\rangle_0~,~
\langle zu\rangle =\lambda \dot{\lambda}\langle z^2\rangle_0~,~
\langle u^2\rangle =\dot{\lambda}^2\langle z^2\rangle_0~.
\end{equation}
We note that Eq.~(\ref{eq:fluid1}) is automatically satisfied by the
ansatz~(\ref{eq:ansatz}). To obtain an
equation for $\lambda$, we integrate Eq.~(\ref{eq:fluid2}) over
$z~dz$. We obtain, for $\lambda$ close to $1$ (small amplitude
oscillations):
\begin{equation}
\ddot{\lambda} + \kappa \dot{\lambda} +\omega^2(\lambda-1) = 0
\end{equation}
with $\kappa$ an effective friction and a breathing oscillation frequency:
\begin{equation}
\omega_{\textrm{br}}=\omega_z\left(3(p-1)+4\right)^{\frac{1}{2}}~.
\label{eq_br1}
\end{equation}
$p$ measures the compression of the cloud:
\begin{equation}
p = \frac{L_{ni}^2}{L_z^2}.
\end{equation}
In experiments where the effective friction is rather small, Eq. (\ref{eq_br1}) is expected to be a fair approximation for the breathing
oscillation frequency. More generally,
assuming a power law two-body interaction force in the gas $1/r^\alpha$,
the simple relation for $\omega_{\textrm{br}}$ in the weak damping
limit becomes~\cite{PhysRevLett.103.224301}:
\begin{equation}
\omega_{\textrm{br}}=\omega_z\left((3-\alpha)(p-1)+4\right)^{\frac{1}{2}}.
\label{eq_br}
\end{equation}
This formula relates $\omega_{\textrm{br}}$ to $\alpha$ and
$p$, and will be used in section~\ref{sec:oscillations}.
Eq.~\eqref{eq_br} was derived in~\cite{PhysRevLett.103.224301} assuming
a velocity independent interaction term, which would be obtained by
linearizing the radiation pressure force (\ref{eq_rad_F}) in velocity.
This is not a reasonable approximation in our experiments
\cite{footnote}, but we
have shown here that \eqref{eq_br} is still expected to provide a
reasonable approximation for the breathing frequency in the limit of
small optical depth.

\section{Experiments}
\label{sec:experiments}

\subsection{Experimental setup}
\begin{figure}
\begin{center}
\includegraphics[scale=0.47]{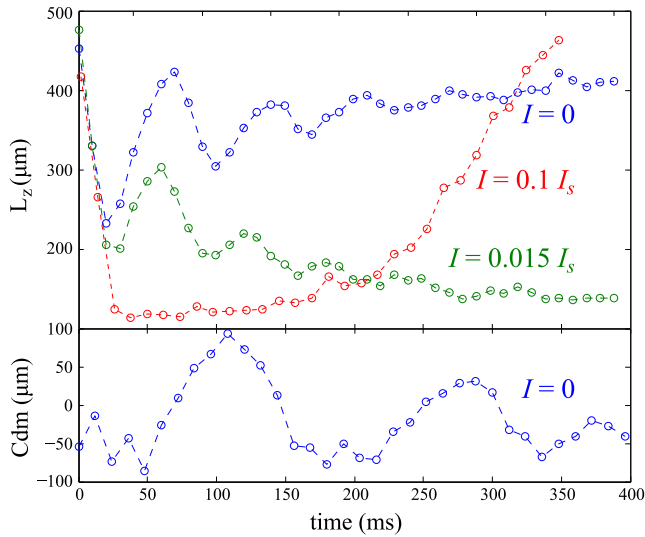}
\caption{\emph{Upper part}: (Color online) Typical temporal evolutions of $L_z$ the \emph{rms} longitudinal size of the atomic cloud for three different 1D beam intensities. The laser detuning is $\delta=-5\Gamma$ for all curves.
\emph{Lower part}: The center of mass (Cdm) position of the atomic gas without the 1D lasers ($I=0$). The $y$ axis origin is arbitrary.}
\label{Temporal}
\end{center}
\end{figure}

The sample preparation is done in the same way as depicted in
\cite{chalony2011}. More details about laser cooling of Strontium in a
magneto-optical trap (MOT) can be found in
\cite{chaneliere2008three}. After laser cooling, around $10^5$ atoms
at $T\simeq 3\,\mu$K are loaded into a far detuned dipole trap made of
a $120$~mW single focused laser beam at $780$~nm. Analysis are
performed using in-situ images taken with a CCD at different instant
of the experimental sequence. The longitudinal profile is obtained
averaging over the irrelevant remaining transverse dimension. We
directly measure the longitudinal trap frequency
$\omega_z=6.7(0.5)$~Hz from relaxation oscillations of the cold cloud
(see example of temporal evolutions in Fig.~\ref{Temporal}). The
radial trap frequency $\omega_\perp=470(80)$Hz is deduced from cloud
size measurements. The beam waist is estimated at $23(2)\,\mu$m
leading to a potential depth of $T_{\textrm{trap}}\simeq 20\,\mu$K.

$50$~ms after loading the dipole trap (corresponding to $t=0$ in
Fig.~\ref{Temporal}), a contra-propagating pair of laser beams,
red-detuned with respect to the $^1\!S_0\rightarrow\,^3\!P_1$
intercombination line at $689$~nm (radiative lifetime: $21\,\mu$s),
is turned on for $400$~ms. These beams, aligned with respect to the
longitudinal axis of the cloud, generate the effective 1D attractive
interaction. When the 1D lasers are on, we apply a $B=0.3$~G magnetic
bias field, for two important reasons: First, the Zeeman degeneracy of
the excited state is lifted such that the lasers interact only with a
two-level system made out of the $m=0\rightarrow m=0$ transition which
is insensitive to the residual magnetic field fluctuation. Second, the
orientation of magnetic field bias, with respect to the linear
polarization of the dipole trap beam, is tuned to cancel the clock (or
transition) shift induced by the dipole trap on the transition of
interest \cite{chalony2011}.

The temperature along the 1D laser beams, in our experimental runs, is
found to be in the range of $1-3\,\mu$K. Even at such low
temperatures, and in sharp contrast with standard broad transitions,
the frequency Doppler broadening $k\bar{v}_z$ remains larger than
$\Gamma$. As a direct consequence, the optical depth $b$ depends on
the exact longitudinal velocity distribution $g(v_z)$ (see
Eqs. (\ref{eq_sigma}) and (\ref{eq_b})) which are not necessarily
gaussian \cite{chalony2011}. Since we measure only the second moment
of the distribution $g(v_z)$, namely: $\bar{v}_z$ or $T$, one has
enough control to assert the $b\ll1$ limit, thus the occurrence of the
self-gravity regime. However we can perform only qualitative tests of
our theory described in section \ref{subsec:stationary_solution}.

At $t=-50$ms, the MOT cooling laser beams are turned off, leaving the
trapped atomic cloud in an out-of-equilibrium macroscopic
state. Without the 1D lasers, we observed a weakly damped oscillation
of the breathing mode and of the center of mass position (blue circles
in Fig. \ref{Temporal}). One notes that damping is caused by
anharmonicity of the dipole trap and not by thermalization of the gas
which is negligible on the experimental timescale. In presence of the
1D laser beams, overdamped or underdamped oscillations of the cloud
are observed.

\subsection{Stationary state's density profile}

\begin{figure}
\begin{center}
\includegraphics[scale=1]{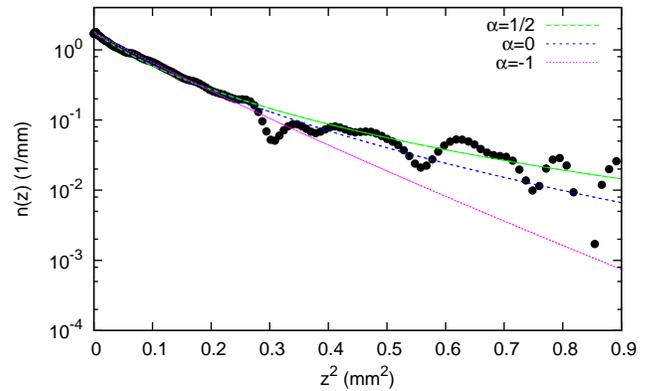}
\caption{(Color online) Density linear distribution for $N=10^5$. The black circles are the
  experimental data with $I=0.02I_s$, $\delta=-6\Gamma$ and $b\simeq
  0.4$. The profiles were symmetrized to improve the signal to noise
  ratio. The curves are least square fits of the data
  using Eq.~(\ref{eq:stat_gen}) containing the
    exact form of the dipole trap and a two-body interaction force
$1/r^\alpha$. The fits are performed for each $\alpha$ by fixing the
  normalization and varying the interaction strength.}
\label{Profile}
\end{center}
\end{figure}

Let us first consider the stationary state in the overdamped situation
(red circles in Fig. \ref{Temporal}). After the transient phase
($t<30$~ms), the \emph{rms} longitudinal size of the atomic gas
reaches a plateau at a minimal value of $L_z\simeq 120\,\mu$m with
$T\simeq 2\,\mu$K. The slow increase of the cloud's size after the
plateau ($t>150$~ms) goes with an increase of the temperature up to
$4\,\mu$K at the end of the time sequence. The origins of the long
time scale evolution are not clearly identified, but it is most likely
due to coupling of the longitudinal axis with the uncooled transverse
dimensions because of imperfect alignment of the 1D laser beams with
the longitudinal axis of the trap and nonlinearities of the trapping
forces. At the plateau where temperature is around $2\,\mu$K the
non-interacting gas is expected to have a \emph{rms} longitudinal size
of $L_z=L_{ni}\simeq 370\,\mu$m. Hence, a clear
compression of the gas by a factor of three is observed. It is due to the attractive
interaction induced by the absorption of the 1D laser beams. Moreover,
the estimated optical depth is $b<0.6$. We then approach the two
previously mentioned conditions --- $b\ll 1$ and $L_z\ll L_{ni}$ ---
for being in the 1D self-gravitating regime as discussed in section \ref{subsec:stationary_solution}. In Fig.~\ref{Profile},
where $b\simeq 0.4$, we test the effective interaction in the gas by
assuming a power law two-body interaction force in the gas
$1/r^\alpha$ and fitting the experimental linear density distribution
for different values of $\alpha$ in the presence of a dipolar trap;
$\alpha=0$ corresponds to 1D gravity. We see that the best fit seems
to be for $\alpha \in[0, 1/2]$.

In absence of the 1D laser beams, we have checked that the
experimental linear density distribution have the expected profile of
a non interacting gas in our dipole trap having a $z_R=1.2(1)$~mm
Rayleigh length.

\subsection{Cloud's longitudinal size}

\begin{figure}
\begin{center}
\includegraphics[scale=0.54]{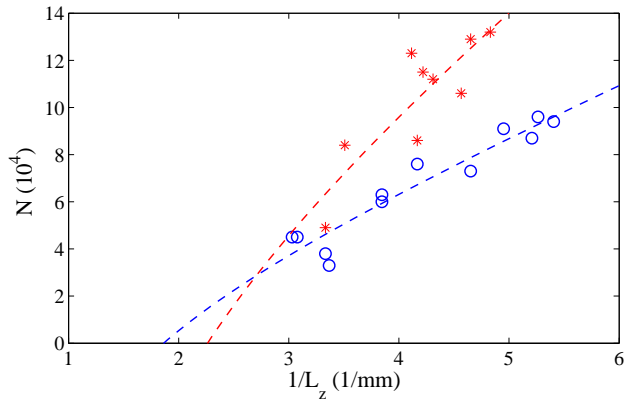}
\caption{(Color online) Dependency of the longitudinal size of the cloud with the
  number of atoms for $\delta=5.7(5)\Gamma$ and $I=0.3I_s$. The blue
  circle (red star) data points correspond to temperature
  $1.5(2)\,\mu$K ($2.1(2)\,\mu$K). The optical depth is in the range
  of 0.6-0.2 according to atoms number variations. The blue and the
  red dashed lines are fits using Eq. (\ref{eq:fitNLz}).}
\label{Nb_Versus_L}
\end{center}
\end{figure}

In the self-gravitating regime a $1/N$ dependency of $L_z$ is expected
at fixed temperature (see \eqref{eq:sech} and the definition of
$L_i$).  Fig.~\ref{Nb_Versus_L} shows that the cloud's size $L_z$ is
in agreement with this prediction for two temperature ranges:
$1.5(2)\,\mu$K (blue circle) and $2.1(2)\,\mu$K (red star). Fits
correspond to the blue dashed line for $1.5(2)\,\mu$K and the red
dashed line for $2.1(2)\,\mu$K. The fitting expression is
\begin{equation}
N=a(1/L_z-L_z/L_{ni}^2)\,,
\label{eq:fitNLz}
\end{equation}
where $a$ and $L_{ni}$ are free parameters depending on the
temperature of the gas. If $L_{ni}\gg L_z$, the self-gravitating
regime is recovered in the fitting expression. However
Eq.~(\ref{eq:fitNLz}) takes into account the presence of a harmonic
trap. Eq.~(\ref{eq:fitNLz}) can be simply derived using the
generalized virial theorem (see Eq.~(11) in
Ref. \cite{werner2008virial}) and it is in perfect agreement with
numerical integrations of Eq.~(\ref{eq:stat}). The fits give
$L_{ni}\simeq 0.5$~mm, slightly larger than the expected value of
$L_{ni}$ at these temperatures. The $1/N$ dependency of $L_z$ in the
self-gravitational regime is consistent with a long-range
interaction with $\alpha=0$. Unfortunately as discussed above, the residual
Doppler effect prevents a quantitative comparison with the prediction
of our model.

\begin{figure}
\begin{center}
\includegraphics[scale=0.63]{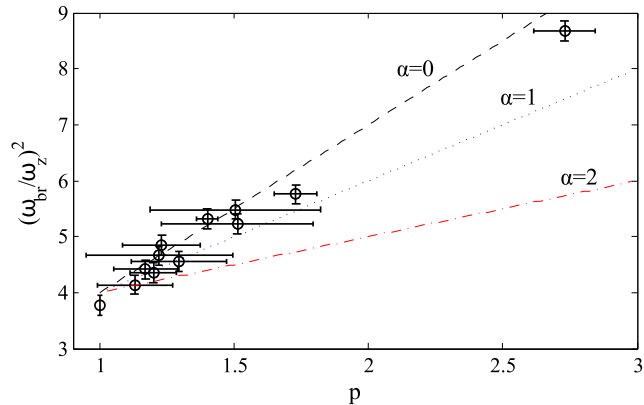}
\caption{(Color online) Comparison for $\alpha=0,1$ and $2$ of the experimental ratio
  $(\omega_{\textrm{br}}/\omega_z)^2$ and the predictions deduced from
  the relation (\ref{eq_br}). The values of $p$ are measured on the
  experiment.}
\label{Oscillation_br}
\end{center}
\end{figure}

\subsection{Breathing oscillations}
\label{sec:oscillations}
Let us now consider the evolution of the trapped cold cloud in the
underdamped situation (as an example see green circles in
Fig. \ref{Temporal}).  Without the 1D lasers, the ratio of the
eigenfrequencies of the breathing mode $\omega_{\textrm{br}}$ and the
center of mass $\omega_z$ is found to be close to two, as expected for
a non-interacting gas in a harmonic trap. As an example the blue
curve, shown in Fig. \ref{Temporal}, gives
$\omega_{\textrm{br}}/\omega_z=1.9(1)$. If now the attractive
long-range interaction is turned on, $\omega_{\textrm{br}}$ is
expected to follow Eq.~\eqref{eq_br} whereas $\omega_z$ should remain
unchanged.

Fig. \ref{Oscillation_br} summarizes the comparisons between the
measured ratio $(\omega_{\textrm{br}}/\omega_z)^2$ and the predictions
deduced from the relation (\ref{eq_br}). $p$ is computed from the
experimental data in the stationary state. We expect $\alpha=0$,
however to judge the nature of the long-range attractive interaction,
three plots respectively for $\alpha=0,~1$ and $2$ are shown. If the
$\alpha=2$ case can be excluded, the experimental uncertainties does
not allow to clearly discriminate between $\alpha=0$ and
$\alpha=1$. In conjunction with Fig.~\ref{Profile}, we conclude that
the system is reasonably well described by a gravitational-like
interaction, $\alpha=0$.

\section{Conclusion and perspectives}

In this paper, we give strong indications of an 1D
gravitational-like interaction in a Strontium cold gas induced by
quasi resonant contra-propagating laser beams. First, we show that in
the self-gravitating limit, the density distribution follows the
theoretically expected profile. Moreover, the scaling of the cloud
size with the number of atoms follows the predicted $1/N$
law. Finally, the modification of breathing frequency of the cloud,
due to the long-range interaction, is correctly described by a
self-gravitating model.

Other phenomena can also be investigated e.g. in relation with plasma physic; Landau damping should be observed studying the return to equilibrium of the system after various perturbations. Moreover, the actual experimental system could be easily extended to 2D geometry suggesting interesting consequences: By
contrast with the 1D case, a 2D self-gravitating fluid undergoes a
collapse at low enough temperature, or strong enough
interaction. Hence, it is conceivable that an experiment similar to
the one presented in this paper, in a pancake geometry, would show
such a collapse \cite{twoDpaper}.

{\bf Acknowledgements:} This work is partially supported by the
ANR-09-JCJC-009401 INTERLOP project. DW wishes to thank
Fr\'{e}d\'{e}ric Chevy for fruitful discussions.


\begin{thebibliography}{13}
\bibitem[{\citenamefont{Olivetti et~al.}(2012)}]{footnote}
  \bibinfo{journal}{For our experimental data points shown in
  Fig.~\ref{Oscillation_br}, $0.3<k\bar{v}_z/|\delta|<1.2$.}
\end{thebibliography}
\end{document}